\documentclass[elsart,usenatbib]{article}
\usepackage{psfig}
\usepackage{amsmath,graphicx}
\usepackage{natbib}
\def\degr{\hbox{$^\circ$}}
\bibpunct{(}{)}{;}{a}{}{,}%to follow the A&A style

\begin{document}

%\headnote{Research Note}
\title{Different Types of Fe $K_\alpha$ Lines from Radiating Annuli near Black Holes}
\author{A.F. Zakharov$^{a,b,c,d}$\footnote{Corresponding author. Tel.:+7 095 1299759; fax:+7 095 8839601.
{\it E-mail address:} zakharov@itep.ru (A.F.~Zakharov).} %\and
~and S.V. Repin$^e$\\
%\thesaurus{02(02.02.1, 02.12.3, 11.19.1, 11.14.1)}
%\offprints{Zakharov A.F., \email{zakharov@itep.ru}} \institute{
$^a$ National Astronomical Observatories of \\
Chinese Academy of Sciences, Beijing 100012, China\\
$^b$Institute of Theoretical and Experimental Physics,\\
           25, B.Cheremushkinskaya st., Moscow, 117259, Russia,\\
%           \email{zakharov@vitep1.itep.ru}
 $^c$Astro Space Centre of Lebedev Physics Institute, 84/32,\\
Profsoyuznaya st.,
             Moscow, 117810, Russia,\\
             %\and
             $^d$Joint Institute for Nuclear Research, Dubna,
             Russia\\
%\and Dipartimento di Fisica Universita di Lecce and INFN, Sezione
%di Lecce,  Italy \and
$^e$Space Research Institute, \\
84/32, Profsoyuznaya st.,
             Moscow, 117810, Russia,\\
%\email{repin@mx.iki.rssi.ru}
}
%\date{Received / accepted }
\maketitle

\begin{abstract}
     Recent X-ray observations of microquasars and Seyfert galaxies
reveal the broad emission lines in their spectra, which can arise
in the innermost parts of accretion disks. A theoretical analysis
of observations and their interpretations were discussed in a
number of papers. We consider a radiating annulus model to
simulate spectral line shapes. That is a natural approximation for
narrow emitting circular rings without extra astrophysical
assumptions about emissivity laws. Recently \cite{Muller03}
presented results of their calculations and classified different
types of spectral line shapes and described their origin. We
clarified their hypothesis about an origin of doubled peaked and
double horned line shapes. Based on results of numerical
simulations we showed that double peaked spectral lines arise
almost for {\it any} locations of narrow emission rings (annuli)
although \cite{Muller03} suggested that such profiles arise for
relatively flat space-times and typical radii for emission region
about $25~r_g$. We showed that triangular spectral lines could
arise for nearest annuli and high inclination angles. We discuss a
possibility of appearance of narrow spectral line shapes as a
result of spiralling evolution of matter along quasi-circular
orbits which could be approximated by narrow annuli.
%\keywords{black hole physics; line: profiles; X-ray galaxies}
\end{abstract}
% \authorrunning{A.F. Zakharov \& S.V. Repin}
% \titlerunning{Different Types of Fe $K_\alpha$ Lines}

\maketitle
\section{Introduction}

More than ten years ago it was predicted that profiles of lines
emitted by AGNs and X-ray binary systems\footnote{Some of them are
microquasars (for details see, for example, \cite{Grein99, Mira00,
Mira02a}).} could have asymmetric double-peaked, double horned or
triangular shape according to \cite{Muller03} classification (e.g.
\cite{Chen89,Fabian89,Robinson90,Dumont90,MPS93}). Generation of
the  broad $K_\alpha$ fluorescence lines as a result of
irradiation of cold accretion disk was discussed by many authors
(see, for example,
\cite{MPP91,MPPS92,MPPS92a,Matt92,MFR93,Bao93,Mart02} and
references therein).
    Recent X-ray observations of Seyfert galaxies, microquasars
and binary systems
(\cite{fabian1,tanaka1,nandra1,nandra2,malizia,paul,sambruna,cui,sulentic1,sulentic2,yaqoob1,yaqoob2,
yaqoob4,ogle1,Miller02,Bianchi04} and references therein) confirm
these considerations in general and reveal broad emission lines in
their spectra with characteristic two-peak profiles. Comprehensive
reviews by \cite{Fabi00,Fabian04,Fabian05} summarize the detailed
discussion of theoretical aspects of possible scenarios for
generation of broad iron lines in AGNs. These lines are assumed to
arise in the innermost parts of the accretion disk, where the
effects of General Relativity (GR) must be taken into account,
otherwise it appears very difficult, if any, to find the natural
explanation of the observed line profile. A formation of shadows
(mirages) is another sample when relativistic effects are very
important
\citep{ZNDI_04,Zakharov_SIGRAV04,Zakharov_Protvino04,ZDIN_AA_05}.

     Numerical simulations of the line structure
could be found in a number of papers, see
\cite{Koji91,Laor91,Bao92,Bao93,BHO94,rauch,Rauch94,bromley,Fan97,pariev2,pariev1,pariev3,
Rusz00,Ma02,Muller03}. They indicate that the accretion disks in
Seyfert galaxies are usually observed at the inclination angle
$\theta$ close to $30\degr$ or less. It occurs because according
to the Seyfert galaxy models, the opaque dusty torque, surrounding
the accretion disk, so, such structure does not allow us to
observe the disk at larger inclination angles.

Recently \cite{Muller03} presented results of their calculations
and classified different types of spectral line shapes and
described their origin. In particular \cite{Muller03} claimed that
usually "... triangular form follows from low inclination
angles...", "...double peaked shape is a consequence of the
space-time that is sufficiently flat. This is theoretically
reproduced by shifting the inner edge to the disk outwards... A
relatively flat space-time is already reached around $25~r_g$..."
We clarified their hypothesis about an origin of doubled peaked
and double horned line shapes. Based on results of numerical
simulations we showed that double peaked spectral lines arise for
{\it almost any} locations of narrow emission rings (annuli)
(except closest orbits as we could see below) although
\cite{Muller03} suggested that such profiles arise for relatively
flat  spacetimes and typical radii for emission region about
$25~r_g$. Using a radiating annulus model we checked the
statements claimed by \cite{Muller03} and clarified it for the
case. We could note here that in the framework of the model we do
not use any assumptions about an emissivity law, but we only
assume that radiating region is a narrow circular ring (annulus).
Thus, below we do not use some specific model on surface
emissivity of accretion (we only assume that the emitting region
is narrow enough). But general statements (which will be described
below) could be generalized on a wide disk case without any
problem.

% In Section $1\degr$ we describe briefly  our model.
% In Section $2\degr$ we discuss  Shakura -- Sunyaev disk model. In Section $3\degr$ we present the results of
% simulations for a narrow annulus model. In Section $4\degr$ we
% discuss results of calculations and present conclusions.

We used an approach which was discussed in details in papers by
\cite{zakh91,zakharov1,zak_rep1,zak_rep2,zak_rep02a,Zak_rep02_Gamma,ZKLR02,zak_rep03_aa,Zak_rep03_AR,ZR_Nuovo_Cim03,Zak_Rep03_Lom,Zakharov_Repin_Ch03,Zak03_Sak,Zak_rep02_xeus,ZR_ASR04,Zakharov_Repin_Pom04,Zak03_Bel,Zak_SPIG04,Zakharov_IJMPA_05}.
The approach was used in particular to simulate   spectral line
shapes. For example, \cite{ZKLR02} used this approach to simulate
an influence of magnetic field on spectral line profiles. This
approach is based on results of qualitative analysis (which was
done by \cite{zakh86,zakh89} for different types of geodesics near
a Kerr black hole). Using first integrals found by \cite{carter}
(see also \cite{wheeler,sharp,chandra}).
 The equations of photon motion in Kerr metric are reduced to
the following system of ordinary differential equations in
dimensionless Boyer -- Lindquist coordinates
(\cite{zakh91,zakharov1,zakharov5,zak_rep1}).

It means that our numerical approach differs from model proposed
by \cite{cunn1,cunn2,karas1,Fan97,rauch,Rauch94}.

\section{Shakura -- Sunayev disk  model}

Here we summarize results of our simulations of spectral line
shapes for Shakura -- Sunyaev model
\citep{zak_rep03_aa,Zak_rep03_AR,Zak_SPIG04}. We analyze the inner
wide part of accretion disk with a temperature distribution which
is chosen according to the \cite{shasun} (see also \cite{LipShak})
with fixed inner and outer radii $r_i$ and $r_o$. Usually a power
law is used for a wide disk emissivity (see, for example,
\cite{Laor91,MPP91,MM96,MKM00,MMK02}) to fit not only spectral
line shapes but X-ray spectra as well. However, another models for
emissivity could not be excluded for so wide class of accreting
black holes, therefore, just to demonstrate how another emissivity
law could change line profiles we exploit  such emissivity law.
Details of spectral line simulations for the Shakura -- Sunyaev
model are described by
\cite{zak_rep03_aa,Zak_rep03_AR,Zak_SPIG04}.

In spite of the fact that \cite{Muller03} noted that double horned
spectral lines arise for power emissivity functions we showed that
also such profiles  are typical for Shakura -- Sunayev model (see
typical spectral line shapes
%in Fig. \ref{A_Muller01} for selected
%position angles $\theta=40\degr, 50\degr, 60\degr$ from left to
%right, see also other spectral line shapes
for the Shakura -- Sunayev model in %presented by
\cite{Zak_rep03_AR,ZR_Nuovo_Cim03,ZR_ASR04,ZR_5SCSLSA} (using the
templates \cite{ZKLR02} calculated distortions of the spectral
line shape by a strong magnetic field, spectral line shapes for
non-flat accretion flows were discussed by \cite{ZMB04}).

     For simulation we assume that
the emitting region lies entirely in the innermost region of
$\alpha$-disk (zone~$a$) from $r_{out} = 10\,r_g$ to $r_{in} =
3\,r_g$ and the emission is monochromatic in the co-moving
frame.\footnote{We use as usual the notation $r_g=2GM/c^2$.}
% The frequency of this emission set as a unity
%by convention.

\section{Results of calculations}

\begin{figure*}[th!]
\begin{center}
\includegraphics[width=\textwidth]{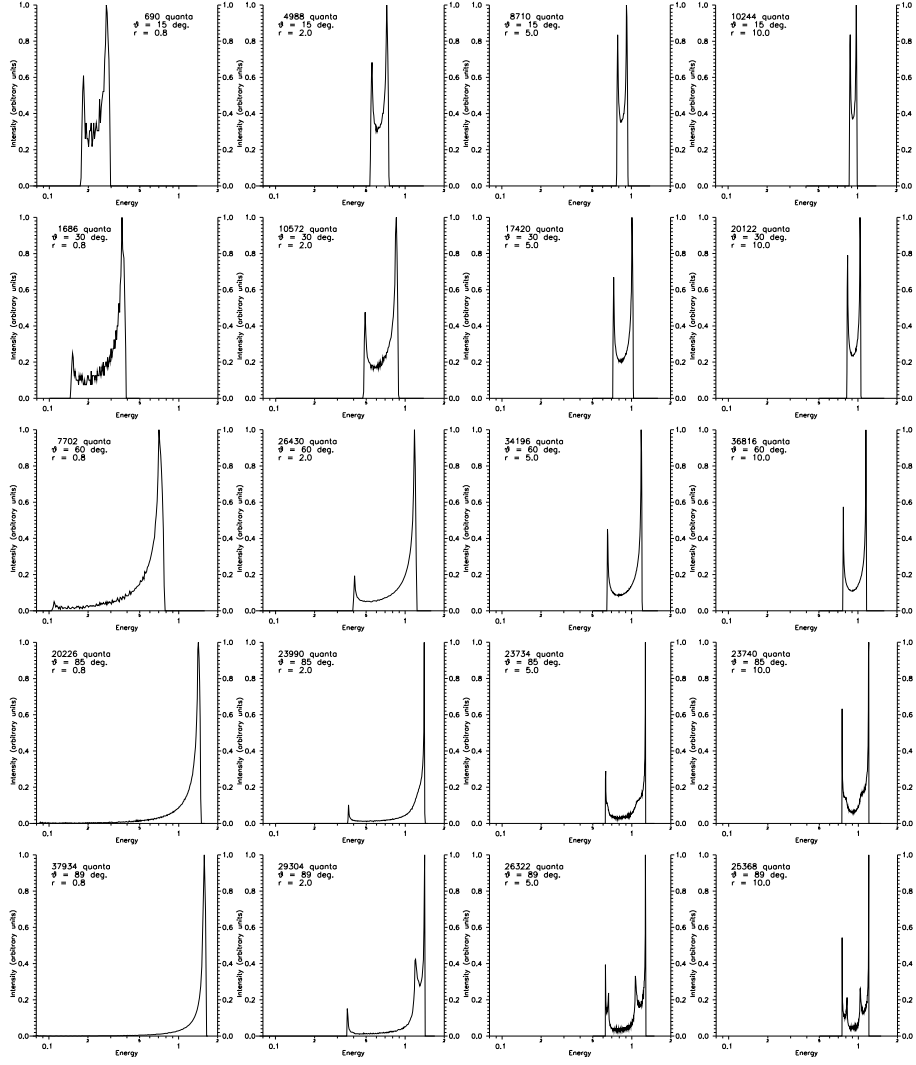}
%\resizebox{\textwidth}{!}{\includegraphics*[36,37][513,600]{sp0710_a.ps}}
\end{center}
  \caption{A set of spectral line shapes for narrow emitting rings (annuli) is shown
  for different short radii and observer position angles.
  Radii take values  0.8,  2, 5, 10 $r_g$(from left to right); angles
  take values $15, 30, 60, 85, 89$ degrees (from top to bottom).}
  \label{A_Muller02}
\end{figure*}

Presenting their classification of different types of spectra line
shapes \cite{Muller03} noted that double peaked shapes arise
usually for emission regions located far enough from black holes.
Earlier, we calculated spectral line shapes for annuli for
selected radii and distant observer position angles and found an
essential fraction of spectral line gallery correspond to double
peaked profiles \cite{zak_rep1}. To check the \cite{Muller03}
hypothesis about an origin of double peaked profiles we calculated
a complete set of spectral line shapes for emitting annuli. Below
we discuss results of our calculations for rapidly rotating black
holes ($a=0.998$).

First, let us assume that an observer is located at the black hole
axis. In this case spectral line shapes look like simulated
$\delta$-function with a redshift corresponding to an annulus
radius, of course the smallest redshift corresponds to the largest
radius \citep{ZR_5SCSLSA}.

If inclination angles are small  ($\theta=15\degr,
\theta=30\degr$), spectral line shapes are double peaked however
one could mention that for radii $r\in (0.7 ,1)$\footnote{Here we
express radius $r$ in $r_g$ units} the intensity of the line is
not very high (see first panels at the first and second rows Fig.
\ref{A_Muller02}) since almost all photons emitted by annuli are
captured by a black hole (see corresponding number of photons
reaching infinity at the selected angle). Since the number of
photons detecting by distant observer is low, there is natural
statistical noise in the spectral line shapes. For larger radii
the statistical noise disappears since a number of photons is much
higher.

For intermediate angle $\theta=45\degr$ and smallest radii a red
peak is very low and it is hardly ever distinguishable from
background \citep{ZR_5SCSLSA}.

For another intermediate angle $\theta=60\degr$ and smallest radii
the spectral line shape has triangular structure, for longer radii
$r=0.8$ and $r=1$ a red peak arises and it is low and probably it
could be distinguished from a background, but for $r=2$ a red peak
is quit clear (see the third row in Fig. \ref{A_Muller02}).

For high inclination angles $\theta > 75\degr$ and small radii
spectral line shapes have a triangular structure (see, for
example, the first panel at the forth row in Fig.
\ref{A_Muller02}). For larger radius $r=2$ and angles $\theta >
75\degr$ spectral line shapes have triangular structure. For
highest inclination angles (for example, $\theta=89\degr$ and
$r=2$) and extra (third) peak arises as it was discussed earlier
by \cite{zak_rep03_aa} (see the second panel in the  fifth row in
Fig. \ref{A_Muller02}).

For larger radii $r > 3$ and angles $\theta = 15\degr, 30\degr,
60\degr,  80\degr$ spectral line shapes have clear double peaked
structure (see, for example, first, second and third, forth and
fifth rows in Fig. \ref{A_Muller02}).

For  radius $r=3$ and angle $\theta = 85\degr$ features of extra
peaks arise (there is a bump) and features are more clear
demonstrated for increasing radii and inclination angles
(approaching the equatorial plane) and the features are clear seen
if radii tend to about $10~r_g$. The effect was discussed earlier
by \cite{zak_rep03_aa} (see also \cite{Dovciak04}).

\section{Discussion and conclusions}

As it was shown in the framework of the simple model the double
peaked spectral line shape arise almost for all parameters $r$ and
$a$ except the case when radii are very small ($r < 2~r_g$) and
inclination angles are in the band $\theta \in [45\degr, 90\degr]$
(for these parameters the spectral line shape has triangular
structure). The phenomenon could be easy understood, since  for
this case the essential fraction of all photons emitted in the
opposite direction in respect to emitting segment of annulus is
captured by a black hole, therefore a red peak is strongly dumped.
For other radii and angles spectral line profiles have double
peaked structure. Therefore we clarify the statement by
\cite{Muller03} that double peaked structure arises if radiation
region is far enough.

If we assume that radiation spot evolved along quasi-circular
orbits from outer to inner radii, then temporal behavior of
profiles could be characterized by a motion from right to left
along each row at each rows in Fig. \ref{A_Muller02}. If we assume
that there is a weak dependence of emissivity function on radius,
then a number of photons characterizes relative intensity in the
line (roughly speaking for $r=0.7$ an intensity (in counts) in 10
times lower then an intensity for $r=2$) therefore in observations
for small radii one could detect only a narrow blue peak but
another part of spectra is non-distinguishable from a background.

One could note also that for fixed radius there is a strong
monotone dependence of intensity on inclination angle (maximal
intensity corresponds to photon motion near equatorial plane and
only a small fraction of ll photons reach a distant observer near
the polar axis). That is a natural consequence of a photon boost
due to a circular motion of emitting fragment of annulus in the
equatorial plane and an influence of spin of a rotating black hole
(see also \cite{Dabrowski97,Miniutti04}).

In the framework of the simple model one could understand that
sometimes the Fe $K_\alpha$ line has only one narrow peak like in
observations of the Seyfert galaxy MCG-6-30-15 by the XMM-Newton
satellite \cite{Wilms01}. If radiating (or illuminating) region is
a narrow annulus evolving along quasi-circular orbits, then
initially two peak structure of the spectral line profile
transform in one peaked (triangular) form. Moreover, an absolute
intensity in the line is increased for smaller radii since a
significant fraction of emitted photons are captured by a black
hole during the evolution of emitting region toward to black hole
in observations we could detect only narrow blue peak and its
height is essentially lower than its height was before for larger
radii. Another part of the triangular spectral line shape could be
non-distinguishable from a background. A relative low intensity
for a triangular spectral line shape could give a narrow single
peak structure in observations.

\subsection*{Acknowledgements}
     AFZ would
like to thank prof.~L.~Piro for fruitful discussions on
observations of narrow Fe $K_\alpha$ lines and their
interpretations %Dipartimento di Fisica Universita di Lecce, INFN,
%Sezione di Lecce for a hospitality (where the final version of the
%paper was prepared)
and profs. F.~DePaolis, G.~Ingrosso, L.~Popovi\'c and Drs.
P.~Jovanovi\'c and A.A.~Nucita for very useful discussions. AFZ is
grateful also to the National Natural Science Foundation of China
(NNSFC)  (Grant \# 10233050) and National Key Basic Research
Foundation Project (Grant \# TG 2000078404) for a partial financial
support of the work.

%    Another author (S.V.R.) is very grateful to
%Prof.~E. Sta\-ro\-sten\-ko, Dr.~A. Salpagarov and Dr.~O. Sumenkova
%for the possibility of fruitful working under this problem. This
%work has been partly supported by Russian Foundation of Basic
%Research, grant 00--02--16108.

Authors are grateful  an anonymous referee for useful remarks and
suggestions.

%\end{acknowledgements}

\end{document}